\documentclass[12pt]{article}
\usepackage{times}
\usepackage{geometry}
\usepackage[utf8]{inputenc}
\usepackage{enumitem,amssymb}
\usepackage{ragged2e}
\newlist{thematic}{itemize}{8}
\setlist[thematic]{label=$\square$}
\usepackage{pifont}
\usepackage{natbib}
\usepackage{sidecap}
\usepackage{graphicx}
\usepackage{floatrow}
\usepackage{multicol}

\begin{document}
\raggedright
\huge
Astro2020 Science White Paper \linebreak

The Demographics and Atmospheres of Giant Planets with the ELTs \linebreak
\normalsize

\noindent \textbf{Thematic Areas:} \hspace*{60pt} $\boxtimes$ Planetary Systems \hspace*{10pt} $\boxtimes$ Star and Planet Formation \hspace*{20pt}\linebreak
$\square$ Formation and Evolution of Compact Objects \hspace*{31pt} $\square$ Cosmology and Fundamental Physics \linebreak
  $\square$  Stars and Stellar Evolution \hspace*{1pt} $\square$ Resolved Stellar Populations and their Environments \hspace*{40pt} \linebreak
  $\square$    Galaxy Evolution   \hspace*{45pt} $\square$             Multi-Messenger Astronomy and Astrophysics \hspace*{65pt} 
  
\begin{multicols}{2}
\textbf{Principal Authors:} \\
Name: Brendan Bowler 
 \linebreak						
Institution: The University of Texas at Austin
 \linebreak
Email: bpbowler@astro.utexas.edu
 \linebreak
Phone: (512) 471-3423 

~\\
 Name: Steph Sallum	
 \linebreak						
Institution: UC Santa Cruz
 \linebreak
Email: ssallum@ucsc.edu
 \linebreak
Phone: (831) 459-4820
\end{multicols}
 
\textbf{Co-authors:} Alan Boss (Carnegie Institution), Timothy Brandt (UC Santa Barbara), Zack Briesemeister (UC Santa Cruz), Marta Bryan (UC Berkeley), Laird Close (U. Arizona), Justin Crepp (University of Notre Dame), Thayne Currie (NASA Ames Research Center), Jonathan Fortney (UC Santa Cruz), Julien Girard (STScI), Rebecca Jensen-Clem (UC Berkeley), Mihkel Kama (University of Cambridge), Adam Kraus (UT Ausitn), Quinn Konopacky (UC San Diego), Michael Liu (University of Hawaii), Mark Marley (NASA Ames Research Center), Christian Marois (NRC-Herzberg), Dimitri Mawet (Caltech), Tiffany Meshkat (IPAC/Caltech), Michael Meyer (University of Michigan), Caroline Morley (UT Austin), Eric Nielsen (KIPAC, Stanford University), Andrew Skemer (UC Santa Cruz), Jason Wang (Caltech), Ya-Lin Wu (UT Austin)
  \linebreak

\textbf{Abstract:}

Gas giants are the most readily detectable exoplanets but fundamental questions about their system architectures, formation, migration, and atmospheres have been unanswerable with the current generation of ground- and space-based facilities. The dominant techniques to detect and characterize giant planets 
--- radial velocities, transits, direct imaging, microlensing, and astrometry --- 
are each isolated to a limited range of planet masses, separations, ages, and temperatures.  These windows into the arrangement and physical properties of giant planets have spawned new questions about the timescale and location of their assembly; the distributions of planet mass and orbital separation at young and old ages; the composition and structure of their atmospheres; and their orbital and rotational angular momentum architectures. 

\smallskip

The ELTs will address these questions by building bridges between these islands of mass, orbital distance, and age.
The angular resolution, collecting area, all-sky coverage, and novel instrumentation suite of these facilities  are needed to provide
a complete map of the orbits and atmospheric evolution of gas giant planets (0.3--10 $M_\mathrm{Jup}$) across space (0.1--100 AU) and time (1 Myr to 10 Gyr). 
This white paper highlights the scientific potential of the GMT and TMT to address these outstanding questions, with a particular focus on the role of direct imaging and spectroscopy 
of large samples of giant planets that will soon be made available with $Gaia$. 

\section{Introduction}
\vskip -.1 in
Characterizing individual exoplanets and surveying their demographics has 
revolutionized our understanding of how planets form and evolve.
Thousands of planets have now been identified using a multitude of complementary detection techniques --- primarily radial velocities, transits, microlensing, and direct imaging. 
Each method adds an important contribution to the broader
story of how micron-sized grains grow by 13--14 orders of magnitude within
a few Myr to produce the diversity of planetary systems we observe today.
These discoveries motivate, guide, and test theories of 
planet formation and evolution spanning five orders of magnitude in mass, separation,
and age.\smallskip

Since the dawn of the exoplanet revolution nearly a quarter century ago, giant planets in particular have exhibited a number of puzzling characteristics 
that in spite of their relative ease of detection have been difficult to theoretically explain, especially in the context of preconceived notions about planetary architectures based on our own solar system.
For instance, hot Jupiters were discovered well within the snow line and with a broad range of eccentricities, implying some planets undergo dramatic dynamical evolution  [1]; 
the orbital planes of many hot Jupiters are misaligned with
respect to the spin axis of their host stars, especially for hot stars [2,3]; 
systems of multiple giant planets can exhibit a diversity of architectures ranging from coplanar orbits (e.g., HR 8799; [4,5]) to strong mutual misalignments (e.g., $\upsilon$ And; [6]);
the radii of many hot Jupiters are inflated (see [7]); 
planetary-mass companions reside at unexpectedly wide separations of hundreds to thousands of AU from their host stars (e.g., [8,9]); 
thick clouds produce unusually red emergent spectra at young ages (e.g., [10,11]); 
and high-altitude hazes are prevalent among close-in irradiated giant planets (e.g., [12]).
These examples 
highlight the continued importance of giant planets in an era increasingly focused on small habitable worlds. \smallskip

Despite great strides in specialized instrumentation, observing strategies, and processing techniques, both technical and astrophysical limitations have prevented the discovery and characterization of giant planets across the entire range of ages and orbital separations.
Instead, each planet detection technique is sensitive to a restricted portion of masses, orbital periods, and ages.  
For example, the transit and radial velocity methods are most sensitive to planets on orbits well within the ice line of Sun-like stars ($\sim$1--3 AU).  These approaches also struggle for young stars, which have fast rotation rates, larger starspot coverage, and increased correlated activity signals on a range of timescales.
Microlensing is most sensitive to the 1--10 AU region around distant field stars, which on average are expected to be old (several Gyr) low-mass M dwarfs (e.g., [13]); moreover, these discoveries cannot be followed up for atmospheric characterization.
Due to contrast limitations, direct imaging with current facilities is most sensitive to self-luminous giant planets orbiting young stars at wide separations (typically $>$10 AU).  
In the next few years, astrometry from $Gaia$ is poised to revolutionize our understanding of giant planet architectures spanning all ages, but only for limited orbital distances of about 1--6 AU. \smallskip

Atmospheric studies of giant planets have had comparable restrictions: transmission spectroscopy and secondary eclipse measurements of gas giants have necessarily focused on short orbital periods and old ages, while direct spectroscopy has largely targeted young planets on wide orbits (see, e.g.,  [14]).
As a result, no systematic survey of giant planet atmospheres and architectures spanning \emph{all} semi-major axes and ages has been possible with existing facilities.
\smallskip

\section{Key Science Questions}{\label{sec:questions}}
\vskip -.1 in
This incomplete census has left open many questions about giant planet formation and evolution: 
\vspace{-5pt}
\begin{itemize}
\setlength\itemsep{0.1em}
\item \textbf{How do giant planets form?} What is the timescale of giant planet assembly? Do planetary accretion processes vary with radial distance from the central star?
\item \textbf{What roles do various molecular ``ice lines" play in giant planet formation?}
\item \textbf{How do planetary orbital and angular momentum architectures evolve over time?} What is the relative importance of different mechanisms of inward or outward migration (disk-planet interactions, planet-planet interactions, or Kozai-Lidov plus eccentric capture)?
\item \textbf{How do non-irradiated giant planet atmospheres evolve over Myr to Gyr timescales?}
What is the mechanism of heavy element enrichment relative to their host stars?  Do the input interior physics and atmospheric chemistry of giant planet evolutionary models match the physical properties and dynamical masses of young and old planets?  
\end{itemize}
\vspace{-15pt}

\section{Bridging the Census of Giant Planets}
\begin{figure}
  \begin{center}
   \includegraphics[width=6.5in]{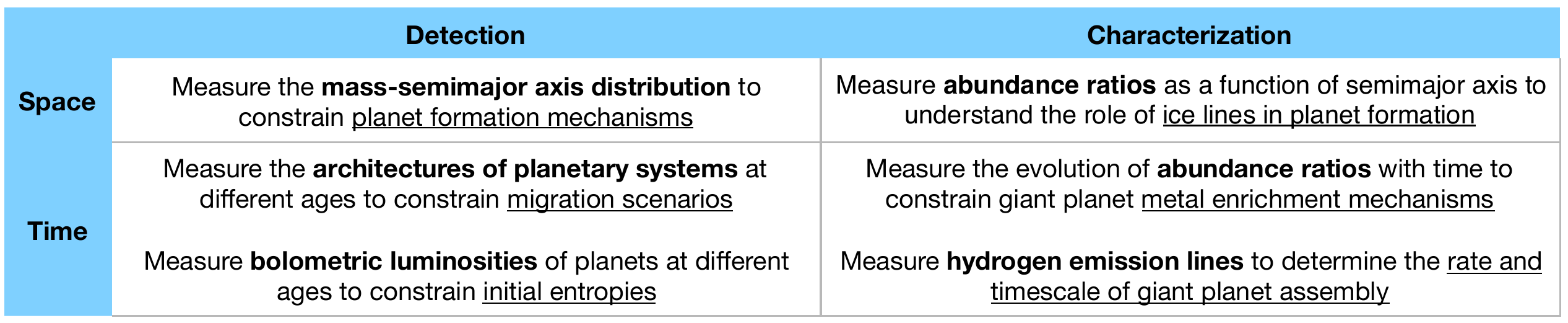}
   \vskip -.2 in
   \caption{\small{Summary of key open questions (underlined) that will be addressed by specific measurements (bold text) to trace the evolution of planet characteristics with semi-major axis (space, top row), and age (time, bottom row).} \label{fig:gpast}}
  \end{center}
 \end{figure}

\vskip -.1 in
The orbits and spectra of giant planets with a range of separations and ages will be powerful tools for understanding how planets form, migrate, and evolve.
In particular, the TMT and GMT have the potential to routinely image older ($>$1 Gyr) and lower-mass ($\approx$0.3--1 $M_\mathrm{Jup}$) planets in thermal emission (3--10 $\mu$m) for the first time (e.g., [15,16]).
Their significant gain in angular resolution will resolve giant planets on $\approx$3--15 AU orbits where formation by core accretion is predicted to be most efficient and observations are revealing a peak in the planet mass function (e.g., [17]).  
Spectroscopic followup will provide an unprecedented view of planetary accretion processes, atmospheric compositions, atmospheric evolution, and rotational angular momentum.
Figure \ref{fig:gpast} summarizes the key observations needed to answer the science questions in Section~\ref{sec:questions}.  
These are discussed in more detail below.
\vspace{-10pt}
 
\subsection{Giant Planet Formation and the Role of Ice Lines}
\vskip -.1 in
The entropy of a planet provides information about the planet assembly process and establishes the future cooling pathway, encapsulated in ``hot-start'' (high-entropy), ``warm-start,'' and ``cold-start'' (low entropy) evolutionary models [18,19,20].
Giant planet luminosities at young ages combined with measurements of their masses can be used to infer the distribution of their entropies as a function of mass and separation.
These initial conditions provide direct information about
energy dissipation during the accretion process and offer a way to distinguish planet formation via the ``bottom-up'' core/pebble accretion scenario versus the ``top-down'' model of disk instability.
\smallskip

Imaging and spectroscopy of hydrogen emission lines
provide constraints on instantaneous mass accretion rates, 
temporal accretion variability, and the overall 
timescale of planet assembly.
The accretion-tracing H$\alpha$ line has already been used to make these measurements for a handful of planets in gapped protoplanetary disks [21,22].
Low-resolution optical/near-infrared spectroscopy of protoplanets also provides the opportunity to distinguish between emission from young planets and forward scattered light by protoplanetary disk material.
High-resolution spectroscopy of these targets can resolve hydrogen emission lines to constrain the accretion geometry from circumplanetary disks (e.g. spherical versus boundary layer versus magnetospheric; see white paper prepared by Sallum et al.).
\smallskip

Medium- and high-resolution spectroscopy of young planets spanning a range of semi-major axes can be used to assess the role of ice lines in giant planet formation and subsequent migration.
C/O ratios of giant planet atmospheres as a population are expected to reflect the location at which they formed relative to prominent snow lines such as $\mathrm{H_2O}$, CO, and $\mathrm{CO_2}$ [23].
Time-dependent dust and chemical evolution models are maturing to a point where predictions for planets can soon be made which will allow for tests of different disk evolution and planet formation processes.
These predictions from chemical modeling of disks 
can be directly tested by searching for trends in C/O atmospheric abundance ratios with semi-major axis.  

\smallskip

These observations will also provide planetary spin measurements through rotationally broadened absorption features [24,25].  The angular momenta of giant planets are primarily set by their accretion histories, which are regulated through interactions with circumplanetary disks.  High-resolution spectroscopy will determine how angular momentum is imparted onto planets from these subdisks and the mechanisms that govern its evolution over time.

\vspace{-10pt}
\subsection{Testing Theories of Migration}
\vskip -.1 in
The distribution of planets as a function of mass, separation, and age encompasses the history of where planets form, how their orbits change over time, and the mechanisms through which this migration occurs. 
Direct imaging at the GMT and TMT, used in conjunction with precision RVs on smaller telescopes to fill in the inner few AU, will provide the most complete census of giant planets spanning 0.1-100 AU.
With a controlled sample of stars this can map the entire mass-period distribution and its evolution with time.
\smallskip

The ability to image Saturn- and Neptune-mass planets at tens to hundreds of AU will help
relate the newly appreciated diversity of protoplanetary disk structure with the diversity of planetary systems.
ALMA in particular has revealed an astounding variety of features in protoplanetary disks --- inner clearings, rings, gaps, asymmetries, and spirals --- which may be signposts of planet formation [26,27]. 
With a complete giant planet census, the architectures of planetary systems can be compared with these disk features to determine the role of giant planets in shaping their formation environments.
Detecting planets within gapped protoplanetary disks, or placing tight upper limits on their masses, will test whether planets cause protoplanetary disk structures.
\smallskip

Long-baseline RV surveys of giant planets have offered important constraints on the radial distribution and migration pathways of
gas giants around field stars (e.g., [28,29]). 
However, these surveys are incomplete at large separations and low masses, and have necessarily focused on old, slowly rotating stars with low activity levels.  As a result, it is unclear \emph{how} and \emph{when} giant planets migrate inward, assuming most form beyond the ice line.
Comparing the planet distribution functions for a range of ages will establish whether most planetary systems undergo orbital migration after formation, and constrain the timescale associated with this process.  
This will directly inform whether disk migration, planet scattering, or Kozai-Lidov oscillations dominate, as the expected timescale for eccentric tidal capture extends well beyond the protoplanetary disk dissipation timescale [30].
Moreover, measurements of the architectures of planetary systems (e.g. orbital spacing and resonances) will also constrain the role of the planetesimal disk in damping planet-planet interactions.

 \begin{figure}
   \includegraphics[width=6.5in]{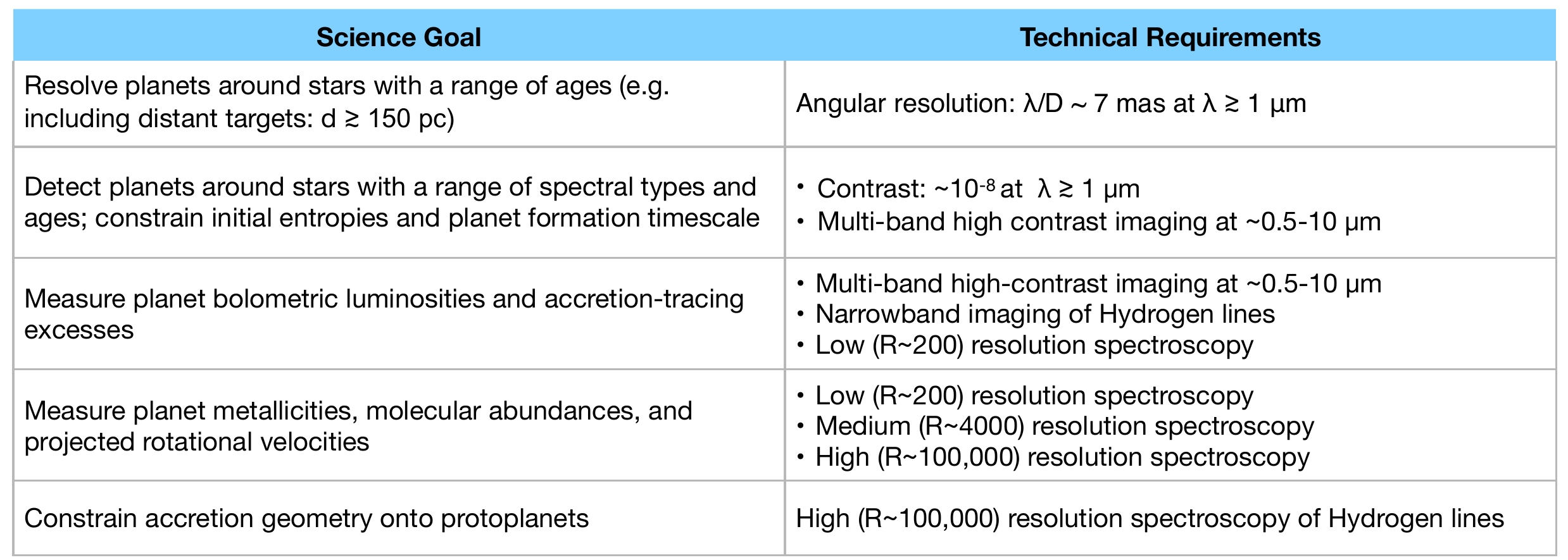}
   \vskip .05 in
   \caption{\small{Science requirements for observing facilities to detect and characterize giant planets around stars with ages spanning Myr to Gyr timescales and separations down to $\sim1$ AU.}\label{fig:scireq}}
   \vskip -.1 in
   \end{figure}
   
\vspace{-10pt}
\subsection{Atmospheric Evolution and Metal Enrichment}
\vskip -.1 in
The atmospheres of the gas and ice giants in our solar system as well as those of a handful of exoplanets have been shown to be enriched in heavy elements (e.g., [31]).
This may be caused by planetesimal bombardment during formation, erosion of a heavy-element core, or accretion from a metal-rich disk.  
These scenarios can be tested by measuring the molecular abundances, compositions, and metallicities of planet atmospheres at spectral resolving powers of several hundred to tens of thousands 
over a broad range of ages and wavelengths, including both the 3--5 $\mu$m region for CO and 8--13 $\mu$m region for NH$_3$ (e.g., [32]).
In particular, observing sub-solar C/H and O/H ratios as well as C/O ratios close to unity would imply intact cores most likely formed via pebble accretion. On the other hand, super-solar C/H and O/H ratios, solar C/O ratios, and an \emph{increasing} atmospheric metallicity over time would point to either core erosion or significant post-formation planetesimal accretion [33].

\vspace{-12pt}
\section{Characterizing \emph{Gaia}-Selected  Giant Planets with the ELTs}
\vskip -.1 in
Unbiased direct imaging surveys of young stars over the past decade have resulted in a steady pace of discoveries, but the overall occurrence rate of super-Jovian planets (5--13~$M_\mathrm{Jup}$) on wide orbits is low compared to planets within a few AU [9,34].  
Fortunately, \emph{Gaia} will provide the community with an age-insensitive sample of tens of thousands of giant planets at orbital separations $\lesssim6$~AU --- a population known to be present based on RV surveys --- by the end of its mission [35].
This sample of ``informed'' targets will remove any potential risk of non-detections and enable detailed imaging and spectroscopy of giant planets spanning the optical to thermal infrared for a broad range of masses, separations, and ages.  These planets will also serve as fundamental calibrators to test 
giant planet cooling models using measurements of  their dynamical masses, effective temperatures, and bolometric luminosities  
(e.g., [36,37]).  \smallskip

\begin{figure}
\floatbox[{\capbeside\thisfloatsetup{capbesideposition={left,center},capbesidewidth=3.0in}}]{figure}[\FBwidth]
{\caption{\small{Properties of $Gaia$ exoplanets that could be imaged by PSI-Red, a potential 2-5 $\mu$m second-generation TMT instrument [16]. This yield assumes a realistic underlying planet population, identifies planets that could be detected by $Gaia$ over a 9-year mission together with WFIRST astrometry (see white paper by Brandt et al.), and determines which of those could be recovered by PSI-Red. The  planet population is identical to that used by the NASA WFIRST team to model the yield capabilities for its upcoming coronagraphic instrument [38]. \label{fig:psired}}}}
{\includegraphics[width=3.in]{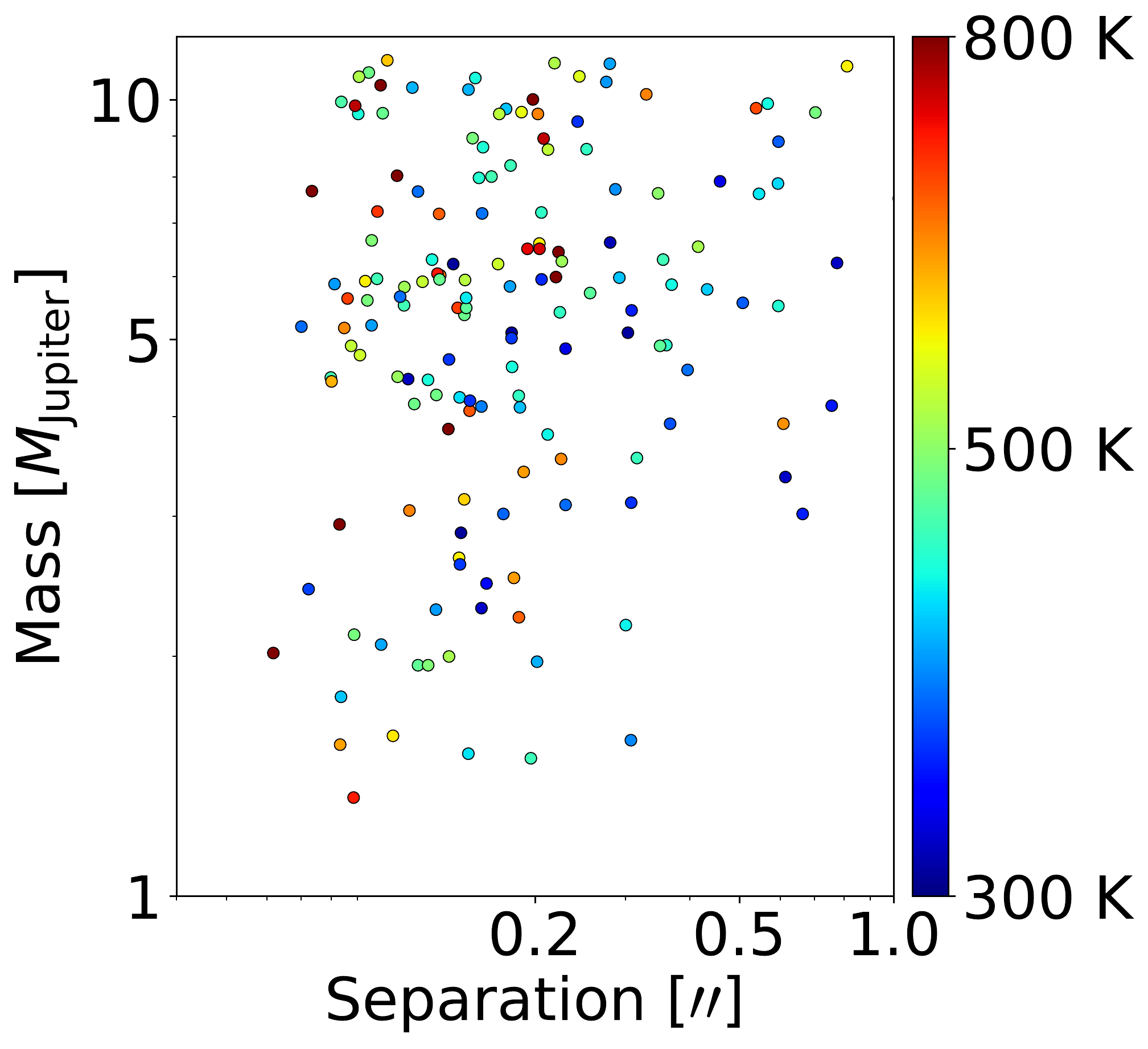}}
\vskip -.3 in
\end{figure}

Imaging and characterizing the \emph{Gaia} planet sample with the goal of addressing the open questions listed in Section~\ref{sec:questions} requires increased sensitivities, higher contrasts, and improved angular resolution compared to current facilities.
To achieve this science, we recommend a robust suite of multiwavelength imaging and spectroscopic instruments on 30 meter-class telescopes.
Current facilities are typically sensitive to young ($<$100 Myr), self-luminous, warm (600-2000 K) super-Jovian planets, whereas the ELTs will be sensitive to older ($>$100 Myr), colder ($T_\mathrm{eff}<$ 600 K), and lower-mass ($<$1 $M_\mathrm{Jup}$) companions --- especially those that will be discovered by $Gaia$ over the next few years.
Figure \ref{fig:scireq} provides an overview of the technical requirements needed to meet these science objectives.
Planned and potential adaptive optics-fed instruments with high-contrast imaging, coronagraphy, and/or spectroscopic capabilities for both the GMT (GMTIFS, GMTNIRS, GMagAO-X, TIGER) and TMT (PSI, MODHIS, IRIS, NIRES, MICHI) are capable of meeting the required contrasts and spectral resolutions for characterizing $Gaia$ planets (see Figure \ref{fig:psired}).  Moreover, these facilities offer long-term multiplexing and upgrade capabilities not possible with space-based telescopes.

\smallskip

\smallskip

Taken together, the instrument and sky coverage complementarity of the GMT and TMT will  
connect the giant planet population with protoplanetary disk structures and evolution;  measure  the  initial  entropies  and  accretion  rates  of  protoplanets;  identify  the  role  of  ice lines and metal enrichment in planet formation and evolution; and determine the dominant mode of giant planet formation and migration.  Altogether these facilities will provide a comprehensive view of giant planet architectures and atmospheres across space ($<$100 AU) and time (1 Myr to 10 Gyr), superseding  our limited  perspective  of  planet  formation  and  evolution  established with the current generation of ground- and space-based telescopes over the past two decades.

\pagebreak
{\Large \textbf{References}}

[1] {Dawson}, R.~I., \& {Johnson}, J.~A. 2018, ARA\&A, 56, 175

[2] Winn, J.~N., Fabrycky, D., Albrecht, S., \& Johnson, J.~A. 2010, ApJ, 718, L145

[3] {Triaud}, A.~H.~M.~J. 2018, {The Rossiter-McLaughlin Effect in Exoplanet
  Research}, 2

[4] Marois, C., Zuckerman, B., Konopacky, Q.~M., Macintosh, B., \& Barman, T. 2010,
  Nature, 468, 1080

[5] Konopacky, Q.~M., Rameau, J., Duchene, G., {et~al.} 2016, ApJL, 829, 1

[6] McArthur, B.~E., Benedict, G.~F., Barnes, R., {et~al.} 2010, ApJ, 715, 1203

[7] {Fortney}, J.~J., \& {Nettelmann}, N. 2010, Space Science Reviews, 152, 423

[8] {Naud}, M.-E., {Artigau}, {\'E}., {Malo}, L., {et~al.} 2014, ApJ, 787, 5

[9] Bowler, B.~P. 2016, PASP, 128, 102001

[10] {Currie}, T., {Burrows}, A., {Itoh}, Y., {et~al.} 2011, ApJ, 729, 128

[11] {Barman}, T.~S., {Macintosh}, B., {Konopacky}, Q.~M., \& {Marois}, C. 2011,
  ApJ, 733, 65

[12]{Sing}, D.~K., {Fortney}, J.~J., {Nikolov}, N., {et~al.} 2016, Nature, 529, 59

[13] {Gaudi}, B.~S. 2012, ARA\&A, 50, 411

[14] Crossfield, I. J.~M. 2015, Publications of the Astronomical Society of the
  Pacific, 127, 941

[15] {Quanz}, S.~P., {Crossfield}, I., {Meyer}, M.~R., {Schmalzl}, E., \& {Held}, J.
  2015, International Journal of Astrobiology, 14, 279

[16] {Skemer}, A.~J., {Stelter}, D., {Mawet}, D., {et~al.} 2018, in Society of
  Photo-Optical Instrumentation Engineers (SPIE) Conference Series, Vol. 10702,
  Ground-based and Airborne Instrumentation for Astronomy VII, 10702A5

[17] {Meyer}, M.~R., {Amara}, A., {Reggiani}, M., \& {Quanz}, S.~P. 2018, A\&A, 612,
  L3

[18] Marley, M.~S., Fortney, J.~J., Hubickyj, O., et al. 2007, ApJ, 655, 541

[19] Fortney, J.~J., Marley, M.~S., Saumon, D., \& Lodders, K. 2008, ApJ, 683, 1104

[20] Marleau, G.~D., \& Cumming, A. 2013, MNRAS, 437, 1378

[21] {Sallum}, S., {Follette}, K.~B., {Eisner}, J.~A., {et~al.} 2015, Nature, 527,
  342
  
[22] {Wagner}, K., {Follete}, K.~B., {Close}, L.~M., {et~al.} 2018, ApJ, 863, L8

[23] {{\"O}berg}, K.~I., {Murray-Clay}, R., \& {Bergin}, E.~A. 2011, ApJ, 743, L16

[24] {Snellen}, I.~A.~G., {Brandl}, B.~R., {de Kok}, R.~J., {et~al.} 2014, Nature,
  509, 63

[25] {Bryan}, M.~L., {Benneke}, B., {Knutson}, H.~A., {Batygin}, K., \& {Bowler},
  B.~P. 2018, Nature Astronomy, 2, 138

[26] {ALMA Partnership}, {Brogan}, C.~L., {P{\'e}rez}, L.~M., {et~al.} 2015, ApJL,
  808, L3

[27] {Andrews}, S.~M., {Huang}, J., {P{\'e}rez}, L.~M., {et~al.} 2018, ApJL, 869,
  L41

[28] {Armitage}, P.~J. 2007, ApJ, 665, 1381

[29] Cumming, A., Butler, R.~P., Marcy, G.~W., {et~al.} 2008, PASP, 120, 531

[30] {Chatterjee}, S., {Ford}, E.~B., {Matsumura}, S., \& {Rasio}, F.~A. 2008, ApJ,
  686, 580

[31] {Crossfield}, I.~J.~M., \& {Kreidberg}, L. 2017, AJ, 154, 261

[32] {Chapman}, J.~W., {Zellem}, R.~T., {Line}, M.~R., {et~al.} 2017, PASP, 129,
  104402

[33] {Madhusudhan}, N., {Bitsch}, B., {Johansen}, A., \& {Eriksson}, L. 2017, MNRAS,
  469, 4102

[34] Galicher, R., Marois, C., Macintosh, B., {et~al.} 2016, A{\&}A, 594, A63

[35] {Perryman}, M., {Hartman}, J., {Bakos}, G.~{\'A}., \& {Lindegren}, L. 2014,
  ApJ, 797, 14

[36] {Bowler}, B.~P., {Dupuy}, T.~J., {Endl}, M., {et~al.} 2018, AJ, 155, 159

[37]  {Brandt}, T.~D., {Dupuy}, T.~J., \& {Bowler}, B.~P. 2018, arXiv e-prints,
  arXiv:1811.07285

[38] {Savransky}, D., \& {Garrett}, D. 2016, Journal of Astronomical Telescopes,
  Instruments, and Systems, 2, 011006

\end{document}